\documentstyle[aps,amsfonts,epsfig,pre,preprint]{revtex} 

\begin{document}

\renewcommand{\theequation}{\arabic{section}.\arabic{equation}} 

\newcommand {\pd}[2]{ \frac{\partial #1}{\partial #2}} \newcommand
{\fd}[2]{ \frac{\delta #1}{\delta #2}} \newcommand {\bv}[1]{
\mathbf{#1} } 

\author{ A.N. Morozov \and A.V. Zvelindovsky \and J.G.E.M. Fraaije} 

\address{Faculty of Mathematics and Natural Sciences, University of
Groningen, Nijenborgh 4, 9747 AG Groningen, The Netherlands} 

\title{Orientational Phase Transitions in the Hexagonal Phase of a Diblock
Copolymer Melt Under Shear Flow} 

\maketitle 

\begin{abstract}
We generalize the earlier theory by Fredrickson [J. Rheol. {\bf 38},
1045 (1994)] to study the orientational behaviour of the hexagonal
phase of diblock copolymer melt subjected to steady shear flow. We use
symmetry arguments to show that the orientational ordering in the
hexagonal phase is a much weaker effect than in the lamellae. We
predict the parallel orientation to be stable at low and the
perpendicular orientation at high shear rates. Our analysis reproduces
the experimental results by Tepe {\it et al.}  [Macromolecules {\bf
28}, 3008 (1995)] and explains the difficulties in experimental
observation of the different orientations in the hexagonal phase. 
\end{abstract}

\section{ Introduction }

Polymeric liquids subjected to shear flow demonstrate a very peculiar
phase behaviour.  Their phase diagrams not only contain regions of
stability of the different symmetry types (lamellae, hexagonally
packed cylinders and so on), but also these regions have an internal
structure.  Application of shear breaks the rotational symmetry
selecting the preferable direction. Thus, at given external parameters
(temperature and shear rate) a certain orientation of the symmetry
pattern with respect to the selected direction is more stable than the
others \cite{main}. 

Experimentally this phenomena was first observed for the lamellar
phase \cite{Koppi}.  In this work a microphase separated
poly(ethylene-propylene)-poly(ethylethylene) diblock copolymer melt
was subjected to an oscillatory shear. Near the order-disorder
transition (ODT) the lamellae have their normal parallel to the
velocity gradient (the parallel orientation) at low shear frequencies,
while at high shear frequencies the lamellae have their normal
parallel to the vorticity direction of the shear flow (the
perpendicular orientation).  At lower temperatures the parallel
orientation is always the most stable one. 

The first theoretical attempt to study the orientational phase
transition was by Cates and Milner \cite{C_M}. They considered the
equation of motion for the order parameter with a coupling of applied
steady shear to the composition fluctuations. They found that flow
completely changes the fluctuation spectrum. As a result, the
fluctuations of the order parameter are suppressed and the ODT
temperature is raised.  The authors suggested that the perpendicular
orientation is stable near the ODT since the composition fluctuations
with wave vectors normal to both the velocity and the velocity
gradient are the least affected by shear. 

Later Fredrickson has shown that the angle-dependence of the fourth
order vertex function is crucial for construction of a realistic
non-equilibrium phase diagram.  Within this framework Fredrickson
reproduced the experimental observation by Koppi {\it et al.} 
\cite{Koppi}. The theory fails to describe the selection of the
parallel orientation at higher frequencies \cite{report}. 

\begin{figure}
\setlength{\unitlength}{1cm} 
\begin{picture}(3,3)
\put(0.0,0){\epsfig{file=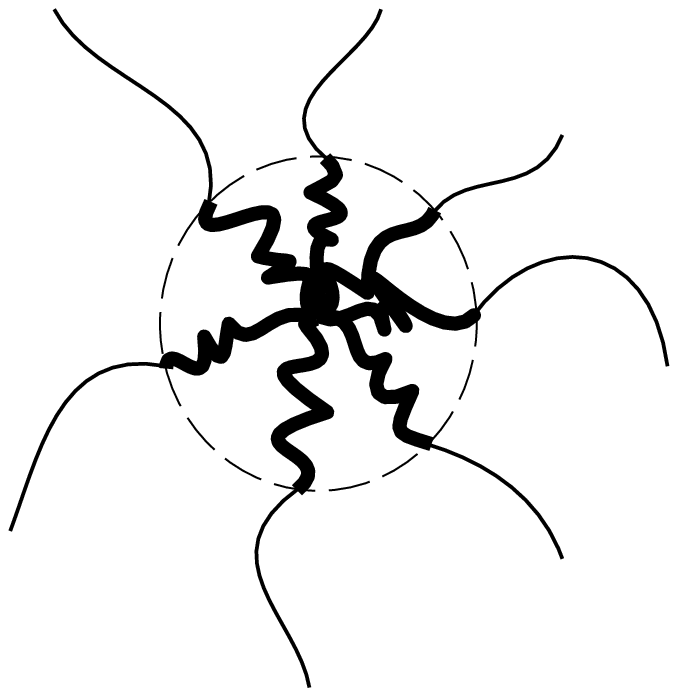,width=3cm}} 
\end{picture}
\caption{ The molecular structure of a cylinder } 
\label{fig1}
\end{figure}

\begin{figure}
\setlength{\unitlength}{1cm} 
\begin{picture}(5,4)
\put(0.0,0){\epsfig{file=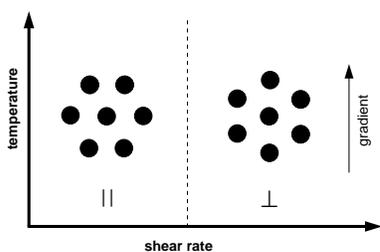,width=5cm}} 
\end{picture}
\caption{ Schematic dynamical phase diagram of hexagonal phase under
shear flow: the parallel ($\phi=0$) and perpendicular ($\phi=\pi/6$)
orientations (see eq.\ref{e2.9}) } 
\label{fig2}
\end{figure}

In the present paper we generalize the Fredrickson theory to describe
the orientational phase transitions in the hexagonal phase of a
diblock copolymer melt subjected to steady simple shear flow.  The
symmetry pattern of this phase consists of hexagonally packed
cylinders made up of blocks of one type immersed in the surroundings
of the other blocks (see Fig.\ref{fig1}).  In equilibrium this phase
appears in between lamellae and body-centered cubic phase
\cite{Leibler,F_H}. The experiments under shear flow show an analogy
with the orientational behaviour of the lamellar phase
\cite{Tepe,Had,Walter}. In slow flow the symmetry pattern has ``2 dots
up'', while in the faster flow the ``1 dot up'' orientation appears
(see Fig.\ref{fig2}).  In order to keep universality we call them the
parallel and the perpendicular orientations, respectively.  There have
been many observations of the ordering of the hexagonal phase
subjected to shear flow \cite{Tepe,Had,Walter}. Recently, Tepe {\it et
al.}  have varied temperature and shear rate and observed both
orientations of the hexagonal pattern in a non-symmetric
polyethylene-poly(ethylenepropylene) (PE-PEP) diblock copolymer
melt. We schematically present their results in Fig.\ref{fig2}. Our
goal is to reproduce this dynamical phase diagram theoretically. 

\begin{figure}
\setlength{\unitlength}{1cm} 
\begin{picture}(3,1.5)
\put(0.0,0){\epsfig{file=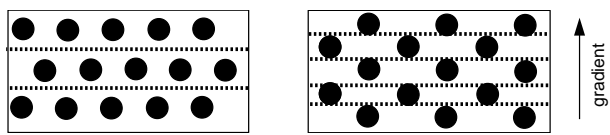,width=8cm}} 
\end{picture}
\caption{ Conventional subdivision of hexagonal pattern in lamellar
layers} 
\label{fig3}
\end{figure}

We start with a speculative analogy with lamellae. Both orientations
of the hexagonal pattern could be considered as parallel lamellae with
different interlamellar distances (see Fig.\ref{fig3}).  High shear
squeezes the pattern and the lamellae with the smallest interlayer
distance are favorable. In very slow shear the intermolecular forces
resulting in microphase separation will play the predominant role and
lamellae with the biggest interlayer distance are stable. Such simple
reasoning already reproduces the main features of Fig.\ref{fig2}. 

We proceed with more exact and motivated analysis.  In Section II we
formulate a dynamical model and study its equilibrium limit.  In
Sections III and IV we apply the methods of Fredrickson \cite{main} to
analyze the high-shear and low-shear behaviour. In the Conclusions we
compare the orientational behaviour of the hexagonal and lamellar
phases and explain the origin of difficulties in experimental study of
this behaviour. 

\section{ Dynamic Equations }
\setcounter{equation}{0}

Let us consider a diblock copolymer melt. The starting point of our
analysis is the dynamic equation for an order parameter
$\psi({\bv{r}})$, which we choose to be a deviation of a local density
of monomers of one block from its average value \cite{Leibler}. In the
present work we use the Fokker-Planck equation for an incompressible
block copolymer melt \cite{main,C_M,Onuki_Kawasaki}: 

\begin{equation}
\label{e2.1}
\pd{P}{t}[\psi,t]=\int_{p}\fd{ }{\psi({\bv{p}})} \left[\mu \left( \fd{
}{\psi(-{\bv{p}})}+ \fd{H}{\psi(-{\bv{p}})} \right)-D\,p_{x}\pd{}{p_y}
\psi({\bv{p}}) \right]P[\psi,t] 
\end{equation}
where $H[\psi]$ is the Landau-Ginzburg Hamiltonian 

\begin{eqnarray}
\label{e2.2}
H[\psi]&=&\frac{1}{2}\int_{q} [\tau
 +(q-q_{0})^2]\psi({\bv{q}})\psi(-{\bv{q}})+\nonumber \\ &
 &\frac{1}{3!}\int_{q_1}\int_{q_2}\int_{q_3}
 \xi({\bv{q}}_1,{\bv{q}}_2,{\bv{q}}_3)
 \psi({\bv{q}}_1)\psi({\bv{q}}_2)\psi({\bv{q}}_3)+ \\ &
 &\frac{1}{4!}\int_{q_1}\int_{q_2}\int_{q_3}\int_{q_4}
 \lambda({\bv{q}}_1, {\bv{q}}_2,{\bv{q}}_3,{\bv{q}}_4)
 \psi({\bv{q}}_1)\psi({\bv{q}}_2)\psi({\bv{q}}_3)\psi({\bv{q}}_4)
 \nonumber 
\end{eqnarray}
and $\mu$ is the Onsager mobility coefficient, which we assume to be
constant (see \cite{main,K_S,Marques} for discussion). 

For a certain configuration of the field $\psi({\bv{q}})$, the
function $P[\psi]$ gives the probability of its realizations. In the
stationary state, $P[\psi]$ reduces to the Boltzmann distribution
$P\sim\exp{(-H[\psi])}$ in the limit $D=0$ \cite{Landau}. 

In the derivation of the equation (\ref{e2.1}) we assumed the standard
flow geometry ${\bv{v}}=D\,y\,{\bv{e}}_x$.  It means that we ignored
any alteration of the velocity profile caused by the internal
structure of the melt. Fredrickson has showed \cite{main} that by
taking into account different viscosities of the two blocks one can
approximate the real velocity profile by the same functional form
${\bv{v}}=D_{\text{eff}}\,y\,{\bv{e}}_x$.  Here $D_{\text{eff}}$ is a
renormalized shear rate: 
\begin{equation}
\label{e2.3}
D_{\text{eff}}=D(1+corrections) 
\end{equation}

In lamellar phase at high shear, the lamellae are perpendicular at
high temperatures, and parallel at low temperatures. Fredrickson has
shown that the orientation-dependent hydrodynamic corrections are
needed to explain this transition.  In hexagonal phase such an effect
at high shear is not known, and we postpone hydrodynamic corrections
to future analysis. Here we approximate $D_{\text{eff}}$ by $D$. 

We take the first and the second cumulants to obtain the equations for 
\begin{eqnarray}
\label{e2.4}
C({\bv{p}})&=&\langle \psi({\bv{p}})\rangle \\ S({\bv{p}})&=&\langle
\psi({\bv{p}})\psi(-{\bv{p}}) \rangle - \langle \psi({\bv{p}})\rangle
\langle \psi(-{\bv{p}})\rangle \nonumber 
\end{eqnarray}
The resulting equations can be greatly simplified if we use the
principle-harmonic approximation for the mean density profile: 
\begin{equation}
\label{e2.5}
C({\bv{p}})=\sum_{i=1}^n a_i
[\delta_{{\bv{p}},q_0{\bv{n}}^{(i)}}+\delta_{{\bv{p}},-q_0{\bv{n}}^{(i)}}]
\end{equation}
where the set of vectors ${\bv{n}}^{(i)}$ determine the lattice
symmetry. The equation (\ref{e2.5}) says that all structures in the
system have the same typical size $\sim1/q_{0}$, where $q_0$
corresponds to the primary peak of the structure factor $S(\bv{p})$
(see the first term in (\ref{e2.2})). 

For the steady state the equations transform to: 
\begin{eqnarray}
\label{e2.6}
h_i&=&\tau a_i +{\frak A}_i+\frac{1}{2}a_i\int_{q}
\lambda(q_0{\bv{n}}^{(i)},-q_0{\bv{n}}^{(i)},{\bv{q}},-{\bv{q}})
S({\bv{q}})+{\frak B}_i \\ 
\label{e2.6.1}
1&=& -\frac{D}{2\mu}p_x \pd{}{p_y}S({\bv{p}})+\nonumber \\ & &
S({\bv{p}}) \left[ \tau+(p-q_0)^2+ \frac{1}{2}\int_{q}
\lambda({\bv{p}},{\bv{q}},-{\bv{p}},-{\bv{q}}) S({\bv{q}}) + {\frak C}
\right] 
\end{eqnarray}
where we defined the structure constants: 
\begin{eqnarray}
\label{e2.7}
{\frak
A}_i&=&\frac{1}{2}\int_{q_1}\int_{q_2}\xi(-q_0{\bv{n}}^{(i)},
{\bv{q}}_1,{\bv{q}}_2)C({\bv{q}}_1)C({\bv{q}}_2)
\nonumber\\ {\frak
B}_i&=&\frac{1}{3!}\int_{q_1}\int_{q_2}\int_{q_3}
\lambda(-q_0{\bv{n}}^{(i)},{\bv{q}}_1,{\bv{q}}_2,{\bv{q}}_3)
C({\bv{q}}_1)C({\bv{q}}_2)C({\bv{q}}_3) \\ {\frak
C}_{\;}&=&\frac{1}{2}\int_{q}\lambda({\bv{p}},{\bv{q}},-{\bv{p}},
-{\bv{q}})C({\bv{q}})C(-{\bv{q}})
\nonumber 
\end{eqnarray}

In equation (\ref{e2.6}) we introduced artificial external fields
$h_i$. This is equivalent to an introduction of the additional term in
the Hamiltonian $H_{ext}=-\int_{q} h({\bv{q}})\psi(-{\bv{q}})$,
$h({\bv{q}})=\sum_{i=1}^n h_i
[\delta_{{\bv{q}},q_0{\bv{n}}^{(i)}}+\delta_{{\bv{q}},-q_0{\bv{n}}^{(i)}}]$,
which describes the interaction with an external field
$h({\bv{q}})$. The fields $h_i$ will allow us to construct a potential
$\Phi$ governing the dynamics. In equilibrium the potential $\Phi$ has
the meaning of the free energy of the system. Thus, by introducing the
fields $h_i$ we obtain an analytic continuation of the free energy to
the dynamic case \cite{main}.  This is possible because our model is
conservative \cite{my_note}. 

As our last simplification we introduce an approximation for the
vertex functions $\xi$ and $\lambda$.  Since we use the
principle-harmonic approximation in (\ref{e2.5}), we assume that all
wavevectors have the same modulus $|{\bv{q}}|=q_0$.  Moreover,
following \cite{Dzyal,KLM,lambda}, we take into account the weak
angle-dependence in the 4th-order vertex function $\lambda$, so 
\begin{eqnarray}
\label{e2.8}
\xi({\bv{q}}_1,{\bv{q}}_2,{\bv{q}}_3)&=&\xi\delta(\hat{{\bv{q}}}_1+
\hat{{\bv{q}}}_2+\hat{{\bv{q}}}_3)
\\ 
\label{e2.8.1}
\lambda({\bv{q}}_1,{\bv{q}}_2,-{\bv{q}}_1,-{\bv{q}}_2)&=&\lambda
\left[1-\beta\left(\hat{{\bv{q}}}_1\cdot\hat{{\bv{q}}}_2 \right)^2
\right] \\ \beta &\ll& 1 \nonumber 
\end{eqnarray}
where $\hat{\bv{q}}={\bv{q}}/q$ denotes the unit vector in the
direction of $\bv{q}$. 

To make the general equations (\ref{e2.6}) and (\ref{e2.6.1}) specific
for the hexagonal phase one needs to calculate the structure constants
(\ref{e2.7}) taking into account the symmetry of the phase. The
average density profile for the hexagonally packed cylinders is given
by (\ref{e2.5}), where the basis vectors ${\bv{n}}^{(i)}$ are: 
\begin{eqnarray}
\label{e2.9}
{\bv{n}}^{(1)}&=&\{0,\cos{\phi},\sin{\phi}\} \nonumber \\
{\bv{n}}^{(2)}&=&\frac{1}{2}\{0,-\cos{\phi}-
\sqrt{3}\sin{\phi},\sqrt{3}\cos{\phi}-\sin{\phi}
\} \\
{\bv{n}}^{(3)}&=&\frac{1}{2}\{0,-\cos{\phi}+
\sqrt{3}\sin{\phi},-\sqrt{3}\cos{\phi}-\sin{\phi}
\} \nonumber 
\end{eqnarray}
The angle $\phi$ defines the orientation of the hexagonal pattern. The
case $\phi=0$ corresponds to the parallel orientation in real space
(see Fig.\ref{fig2}). 

Using (\ref{e2.7}), (\ref{e2.9}) and (\ref{e2.8}), (\ref{e2.8.1}) we
obtain for the structure constants: 
\begin{eqnarray}
\label{str_const}
{\frak A}_i&=&\xi a_k a_l \nonumber \\ {\frak
B}_i&=&a_i\lambda\left[a_k^2 \left(1-\frac{\beta}{4}\right)+a_l^2
\left(1-\frac{\beta}{4}\right)+\frac{1}{2}a_i^2
\left(1-\beta\right)\right] \\ {\frak C}_{\;}&=&\lambda
\left[a_1^2\left(1-\beta({\hat{\bv{p}}}\cdot{\bv{n}}^{(1)})^2
\right)+a_2^2\left(1-\beta({\hat{\bv{p}}}\cdot{\bv{n}}^{(2)})^2
\right)+a_3^2\left(1-\beta({\hat{\bv{p}}}\cdot{\bv{n}}^{(3)})^2
\right) \right] \nonumber \\ &~&~~~~~~~~~~~~~~~~~~~~~~~~~i\neq k\neq l
\nonumber 
\end{eqnarray}
We introduce the notation: 
\begin{eqnarray}
\label{e2.10}
\sigma(\hat{{\bv{p}}})&=&\frac{\lambda}{2}\int_q
S({\bv{q}})\left[1-\beta(\hat{{\bv{p}}}\cdot\hat{{\bv{q}}})^2 \right]
\\ 
\label{e2.10.1}
r-\hat{\bv{p}}\cdot{\bv{\stackrel{\leftrightarrow}{e}}}
\cdot\hat{{\bv{p}}}&=&\tau+\sigma(\hat{{\bv{p}}})+\lambda
\sum_{i=1}^3 a_i^2 \left[1-\beta(\hat{{\bv{p}}}\cdot{\bv{n}}^{(i)})^2
\right] \\ 
\label{e2.10.2}
S_0({\bv{p}})&=&\left[r-\hat{\bv{p}}
\cdot{\bv{\stackrel{\leftrightarrow}{e}}}\cdot\hat{{\bv{p}}}+(p-q_0)^2
\right]^{-1} 
\end{eqnarray}
and rewrite the steady-state equations in the final form: 
\begin{eqnarray}
\label{e2.11}
-\frac{D}{2\mu}p_x\frac{\partial}{\partial
p_y}S({\bv{p}})&+&S({\bv{p}})S_0^{-1}({\bv{p}})=1 \\ 
\label{e2.12}
h_i=(r-\hat{\bv{n}}^{(i)}\cdot{\bv{\stackrel{\leftrightarrow}{e}}}
\cdot\hat{{\bv{n}}}^{(i)})a_i&+&\xi
a_k a_l-\frac{1}{2}\lambda a_i^3 (1-\beta)\\ i\neq &\!k&\neq l
\nonumber 
\end{eqnarray}

Notation (\ref{e2.10})-(\ref{e2.10.2}) has a clear physical
meaning. The fluctuation integral (\ref{e2.10}) takes into account the
fluctuations of the order parameter \cite{Braz} and renormalizes the
temperature in the system.  Because of the angle-dependence of the
4th-order vertex function $\lambda$ (eq.\ref{e2.8.1}), the
renormalized temperature also has an angle-dependence. Expanding it to
the first order in $\beta$ (one can easily check that the anisotropy
tensor $e_{ij}$ is of order $O(\beta)$), we extract this
angle-dependence and get (\ref{e2.10.1}), where $r$ denotes the
$\bv{p}$-independent part of the renormalized temperature and
$-\hat{\bv{p}}\cdot{\bv{\stackrel{\leftrightarrow}{e}}}\cdot\hat{{\bv{p}}}$
adsorbs the other terms.  Finally, $S_0({\bv{p}})$ is an equilibrium
structure factor, which in the limit $\beta=0$ reduces to the one
studied by Brazovskii \cite{Braz}, Fredrickson and Helfand \cite{F_H}. 

\subsection*{Equilibrium analysis}

The equations (\ref{e2.11}) and (\ref{e2.12}) describe the behaviour
of the hexagonal phase at any shear rate.  For the particular case
($D=0$, $\beta=0$) these equations were studied in a number of
articles \cite{F_H,Marques,Braz}. Before studying the dynamics we want
to show the influence of the angle-dependence in $\lambda$
(\ref{e2.8.1}) on the equilibrium phase behaviour, so we construct the
free energy at equilibrium $D=0$. The equation (\ref{e2.11}) reads: 
\begin{equation}
S({\bv{p}})=S_0({\bv{p}})=\left[r-\hat{\bv{p}}
\cdot{\bv{\stackrel{\leftrightarrow}{e}}}\cdot\hat{{\bv{p}}}+(p-q_0)^2
\right]^{-1} 
\end{equation}
Performing the integration in (\ref{e2.10}) up to the first order in
$\beta$ we get: 
\begin{equation}
\label{sigma_eq}
\sigma(\hat{{\bv{p}}})\equiv\sigma_{eq}=\frac{\alpha\lambda}
{\sqrt{r}}\left[1-\frac{1}{3}\beta
(\hat{\bv{p}}\cdot\hat{\bv{p}})+\frac{e_{ii}}{6r} \right] 
\end{equation}
Here $e_{ii}$ denotes the trace of the anisotropy tensor
${\bv{\stackrel{\leftrightarrow}{e}}}$.  Separating
$\hat{\bv{p}}$-dependent terms in $\sigma(\hat{{\bv{p}}})$ with the
help of (\ref{e2.10.1}) we obtain: 
\begin{eqnarray}
\label{r_eq}
r&=&\tau+\frac{\alpha\lambda}{\sqrt{r}}+\frac{\alpha\lambda}
{6\,r^{3/2}}e_{ii}+\lambda(a_1^2+a_2^2+a_3^2)
\\ 
\label{e_eq}
e_{ij}&=&\frac{\alpha\lambda}{\sqrt{r}}\frac{\beta}{3}\delta_{ij}+
\beta\lambda[a_1^2\,n_{i}^{(1)}\,n_{j}^{(1)}+a_2^2\,n_{i}^{(2)}
\,n_{j}^{(2)}+a_3^2\,n_{i}^{(3)}\,n_{j}^{(3)}]
\end{eqnarray}
The equations of motion for $a_i$ (\ref{e2.12}) have a potential form: 
\begin{equation}
\label{pot}
h_i=\frac{1}{2}\frac{\partial\Phi}{\partial a_i} 
\end{equation}
To integrate (\ref{pot}) one needs to take into account dependence of
$r$ on $a_i$ and treat an integral of a composite function: 
\begin{equation}
\label{r_composite}
\int_{0}^{a_i}da_{i}(\dots)=\int_{r0}^{r}dr\frac{\partial
a_i}{\partial r}(\dots) 
\end{equation}
where the Jacobian $\frac{\partial a_i}{\partial r}$ can be found from
(\ref{r_eq}), and $r_0$ is a temperature at which the disordered phase
loses stability, given by equation (\ref{r_eq}) with
$a_1=a_2=a_3=0$. Integrating (\ref{pot}) we obtain the free energy: 
\begin{eqnarray}
&\Phi&=\Phi'+\Phi_{str} \\
&\Phi&'=\frac{r^2-r_0^2}{2\lambda}+\alpha(1-\frac{\beta}{2})(\sqrt{r}-
\sqrt{r_0})+\frac{1}{2}\alpha\tau\beta(\frac{1}{\sqrt{r}}-
\frac{1}{\sqrt{r_0}})+\frac{1}{6}\alpha^2
\lambda \beta(\frac{1}{r}-\frac{1}{r_0})\\
&\Phi&_{str}=-\frac{1}{4}\lambda(1+\beta)(a_1^4+a_2^4+a_3^4)+2\xi a_1
a_2 a_3-\frac{1}{4}\beta \lambda \left(a_1^2 a_2^2+a_1^2 a_3^2+a_2^2
a_3^2 \right) 
\end{eqnarray}
where we have separated terms determined by the structure constants. 

The amplitudes $a_i$ that minimize the potential $\Phi$ are found by
solving the set of equations $h_i=0$. Its most important feature is
that $\sigma$ does not depend on the ${\bv{n}}^{(i)}$. Thus, the
resulting equations are isotropic with respect to $a_i$ and the
solution has the form $a_1=a_2=a_3={\frak a}$. 

\begin{figure}
\setlength{\unitlength}{1cm} 
\begin{picture}(10,8)
\put(-0.5,0){\epsfig{file=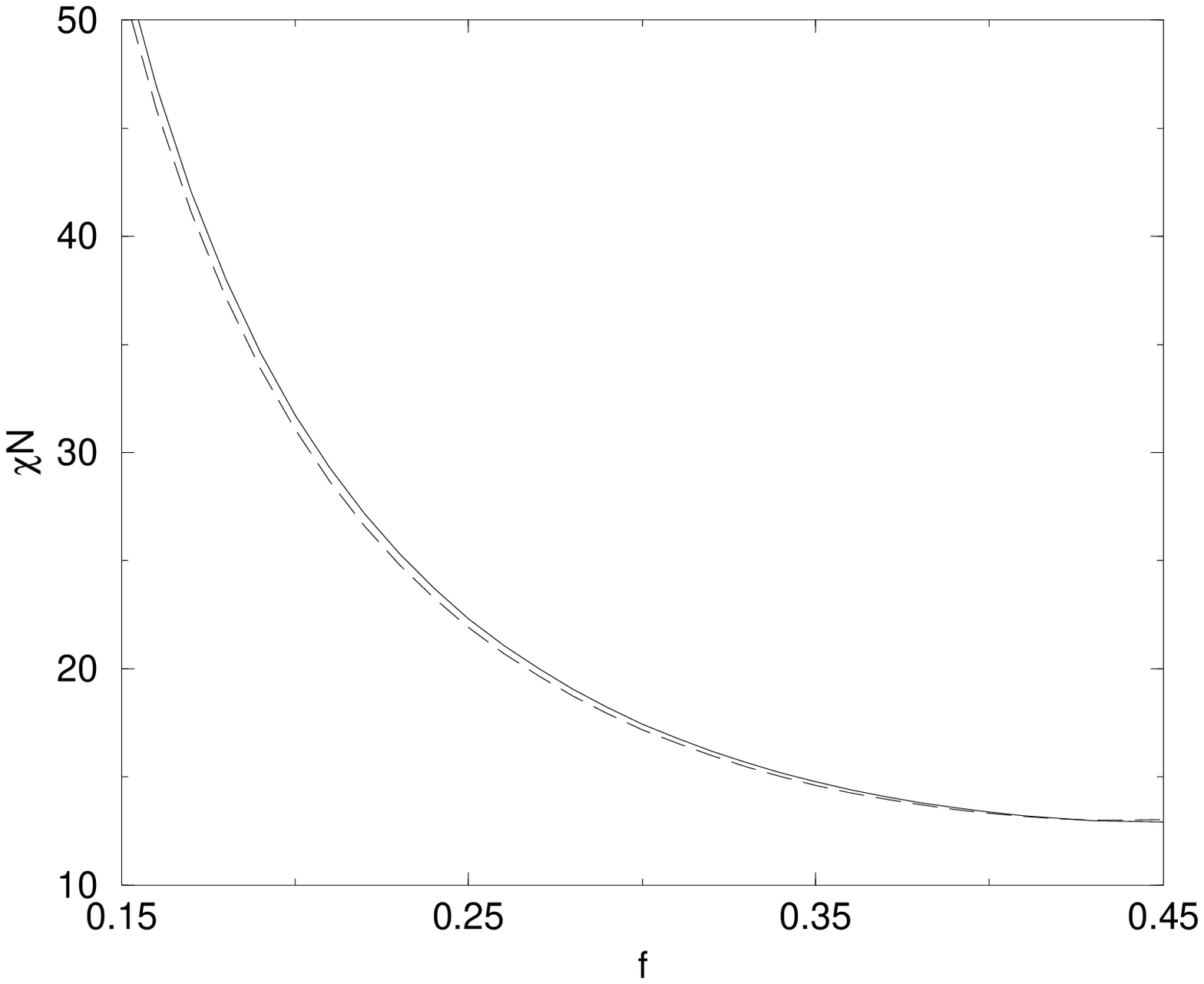,width=8cm}} 
\end{picture}
\caption{ The spinodal temperature ({\it solid line: $\beta=0$, dashed
line: $\beta=0.5$}) } 
\label{graph1}
\end{figure}

\begin{figure}
\setlength{\unitlength}{1cm} 
\begin{picture}(10,8)
\put(-0.5,0){\epsfig{file=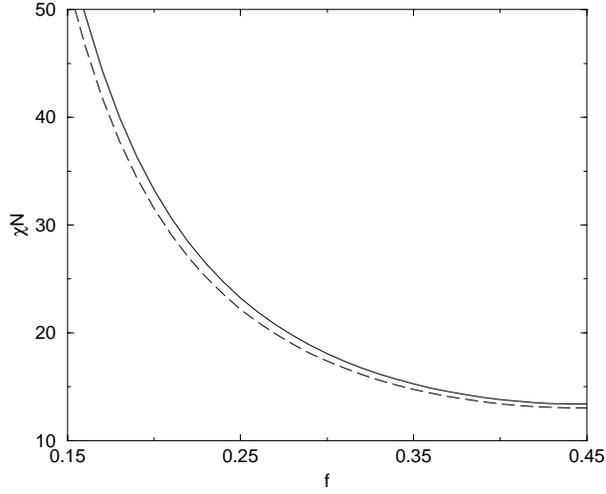,width=8cm}} 
\end{picture}
\caption{ The transition temperature between lamellae and cylinders
({\it solid line: $\beta=0$, dashed line: $\beta=0.5$}) } 
\label{graph2}
\end{figure}

To explore the stability region of the hexagonal phase we need to
solve the resulting system of equations for given temperature and
composition: 
\begin{eqnarray}
\label{r_equation}
&r&=\tau+\frac{\alpha\lambda}{\sqrt{r}}+\frac{(\alpha\lambda)^2}{6
r^2}\beta+\frac{\alpha \lambda^2}{2r^{3/2}}{\frak
a}^2\beta+3\lambda{\frak a}^2 \\
&r_0&=\tau+\frac{\alpha\lambda}{\sqrt{r_0}}+\frac{(\alpha\lambda)^2}{6
r_0^2}\beta \\ 
\label{a_eq}
&r&-\frac{1}{3}\beta\frac{\alpha\lambda}{\sqrt{r}}-\lambda {\frak a}^2
(\frac{1}{2}+\beta)+\xi {\frak a}=0 \\
&\Phi&=\Phi'-\frac{3}{4}\lambda{\frak a}^4(1+2\beta)+2\xi{\frak a}^3 
\end{eqnarray}
The calculation technique of Fredrickson and Helfand \cite{F_H} allows
us to demonstrate the influence of the $\beta$-terms on the
equilibrium phase diagram. In Fig.\ref{graph1} and Fig.\ref{graph2} we
show that the approximation (\ref{e2.8.1}) results in a
non-significant quantitative shift of values. The effect of $\beta$ on
the equilibrium phase diagram is tiny, but in the next section we will
demonstrate that $\beta$ is dominant in explaining the dynamical
orientational ordering.

\section{Strong-shear behaviour}
\setcounter{equation}{0} 

To study the orientational dynamics in the strong-shear regime we need
to solve (\ref{e2.11}) and (\ref{e2.12}) in the limit
$D\rightarrow\infty$.  In this case it is impossible to obtain a
solution of (\ref{e2.11}) as a perturbation series in $1/D$. Instead,
we use approach developed by Cates and Milner \cite{C_M}. One can
apply the RG methods \cite{Onuki_Kawasaki} to find the asymptotic
behaviour of the structure factor
$S({\bv{p}},D\rightarrow\infty)=S_{\infty}({\bv{p}})$.  Then, at large
shear rates the structure factor can be approximated by interpolating
between $S_0({\bv{p}})$ and $S_{\infty}({\bv{p}})$. Cates and Milner
introduced the following approximation: 
\begin{equation}
S^{-1}({\bv{p}})=S_0^{-1}({\bv{p}})+S_{\infty}^{-1}({\bv{p}})=
{r-{\bv{p}}\cdot{\bv{\stackrel{\leftrightarrow}{e}}}\cdot{\bv{p}}+
(p-q_0)^2+\frac{1}{c_0}\left(
\frac{D |p_x p_y|}{\mu \, \alpha^{1/2}}\right)^{2/3}} 
\end{equation}
where $c_0=\frac{1}{3}(48\pi)^{1/3}\Gamma\left(\frac{1}{3}\right)$. To
the leading order in $D$ the fluctuation integral is equal to
\cite{main}: 
\begin{equation}
\label{sigma_high}
\sigma({\bv{\hat{p}}})=(\alpha\lambda)^{2/3}{\frak{D}}
[I_1-\beta(I_2{\hat{p}_x}^2+I_2{\hat{p}_y}^2+I_3{\hat{p}_z}^2) ] 
\end{equation}
where 
$$
{\frak{D}} =
\frac{\sqrt{c_0}}{(48\pi)^{1/3}}\left(\frac{D^*}{D}\right)^{1/3}\;,\;
D^*=\mu\lambda\sqrt{\alpha}
$$
and 
$$
I_{1}=\frac{\Gamma\left(\frac{1}{2}\right)\Gamma\left(
\frac{1}{3}\right)^2}{2\pi\Gamma\left(\frac{7}{6}\right)}\approx2.2
\;,\;
I_{2}=\frac{\Gamma\left(\frac{1}{2}\right)\Gamma
\left(\frac{1}{3}\right)
\Gamma\left(\frac{4}{3}\right)}{2\pi\Gamma\left(
\frac{13}{6}\right)}\approx0.6
\;,\;
I_{3}=\frac{\Gamma\left(\frac{3}{2}\right)\Gamma
\left(\frac{1}{3}\right)^2}{2\pi\Gamma\left(\frac{13}{6}
\right)}\approx0.9
$$
For the external fields $h_i$ the equation (\ref{e2.12}) results in: 
\begin{equation}
\label{h_high}
h_i=\left[\tau+\sigma({\bv{n}}^{(i)})+\lambda A
\left(1-\frac{1}{4}\beta \right)\right]a_{i}+\xi
a_{\alpha}a_{\beta}-\frac{1}{2}\lambda a_{i}^3
\left(1+\frac{1}{2}\beta\right) 
\end{equation}
where 
$$
A=a_1^{2}+a_2^2+a_3^2 
$$
The potential $\Phi$ is given by (\ref{pot}). Unlike the equilibrium
situation we do not have any problem with integration ((\ref{h_high})
does not depend on both $r$ and $e_{ij}$), and the equation for the
free energy is straightforward: 
\begin{equation}
\label{fe_high}
\Phi=\tau A +
\sum_{j=1}^{3}\sigma({\bv{n}}^{(j)})a_j^2+\frac{1}{2}\lambda
\left(1-\frac{1}{4}\beta \right) A^2+2\xi a_1 a_2
a_3-\frac{1}{4}\lambda\left(1+\frac{1}{2}\beta
\right)\sum_{j=1}^{3}a_j^4 
\end{equation}
The amplitudes $a_i$ are the solutions of the equations 
\begin{equation}
\label{h=0}
h_i=0 
\end{equation}
In the absence of the term $\sigma({\bv{n}}^{(j)})a_{j}^2$ these
equations have a uniform solution $a_1=a_2=a_3$.  The presence of this
term breaks the symmetry of the equations and allows for the nonequal
amplitudes. We construct the solutions of (\ref{h=0}) as a
perturbation series in two small parameters: $\frak D$ and
$\beta$.  The highest order in the perturbation series for the
amplitudes is determined by the following argument. Our goal is to
obtain an angle-dependent free energy. This angle-dependence can only
appear via various but symmetric combinations of the basis vectors
${\bv{n}}^{(i)}$. Let us discuss an example: 
$$
n^{(1)\,2x}_y+n^{(2)\,2x}_y+n^{(3)\,2x}_y 
$$
This combination is angle-dependent if $x\geq3$.  One can easily check
that it is also true for the other symmetric combinations. The only
way for such a combination to enter the equation for the free energy
is via a term like $\sigma^3$, which is proportional to $\beta^3{\frak
D}^3$.  Thus, in what follows we keep terms up to $O(\beta^3{\frak
D}^3)$. Solving (\ref{h_high}) for the amplitudes and substituting
them into (\ref{fe_high}), we obtain 
\begin{equation}
\Phi=\Phi_0+\Phi_1 {\frak{D}} + \Phi_2 {\frak{D}}^2 + \Phi_3
{\frak{D}}^3 
\end{equation}
The coefficients $\Phi_i$ are given by: 
\begin{eqnarray}
\Phi_0&=&\frac{-2+15(2-5x)x-2(1-10x)^{3/2}}{125}\frac{\xi^4}{\lambda^3}
\nonumber \\ \Phi_1&=& \frac{3}{25}I_1 \left(1+\sqrt{1-10x} \right)^2
\frac{\xi^2 \alpha^{2/3}}{\lambda^{4/3}}\nonumber \\
\Phi_2&=&-\frac{3}{5}I_1^2(1+\frac{1}{\sqrt{1-10x}})\left(\alpha^4
\lambda \right)^{1/3} \nonumber \\ \Phi_3&=&
\left[-\frac{I_1^3}{(1-10x)^{3/2}}+(I_3-I_2)^3 \beta^3
\frac{\sqrt{1-10x}(122-5x)+27(14-5x)}{16 (8+x)^3}\cos{6\phi}
\right]\left(\frac{\alpha \lambda}{\xi}\right)^2 \nonumber 
\end{eqnarray}
Here we have introduced the dimensionless temperature
$x=\tau\lambda/\xi^2$ and kept only the leading terms in $\beta$ for a
given order in $\frak D$.  The second term in $\Phi_3$ is the
angle-dependent contribution we were looking for. 

The spinodal temperature is obtained from the condition:
$\Phi\mid_{\tau=\tau_s}=0$. It reads: 
\begin{equation}
\tau_s=\tau_s^{(0)}+\tau_s^{(1)} {\frak{D}} + \tau_s^{(2)}
{\frak{D}}^2 + \tau_s^{(3)} {\frak{D}}^3 
\end{equation}
where: 
\begin{eqnarray}
\tau_s^{(0)}&=&\frac{4}{9(5-2\beta)}\frac{\xi^2}{\lambda} \nonumber \\
\tau_s^{(1)}&=&-\left[I_1-\frac{1}{2}\beta\left(I_2+I_3
\right)\right](\alpha\lambda)^{2/3} \nonumber \\
\tau_s^{(2)}&=&\frac{9\beta^2 (5-2\beta)^2 (I_3-I_2)^2}{64
(13-7\beta)}\frac{\alpha^{4/3}\lambda^{7/3}}{\xi^2} \nonumber \\
\tau_s^{(3)}&=&-\frac{81(I_3-I_2)^3 \beta^3 (5-2\beta)^5}{512
(13-7\beta)^3}\left(\frac{\alpha\lambda^2}{\xi^2}\right)^{2}\cos{6\phi}
\nonumber 
\end{eqnarray}

Finally, we calculate the transition temperature between cylinders and
lamellae. Up to the first order in ${\frak D}$ we obtain: 
\begin{equation}
\tau_{tr}=-\frac{7+3\sqrt{6}}{5}\frac{\xi^2}{\lambda}+
I_1(7+3\sqrt{6})(\alpha\lambda)^{2/3}{\frak
D} 
\end{equation}

\begin{figure}
\setlength{\unitlength}{1cm} 
\begin{picture}(5,4)
\put(0.0,0){\epsfig{file=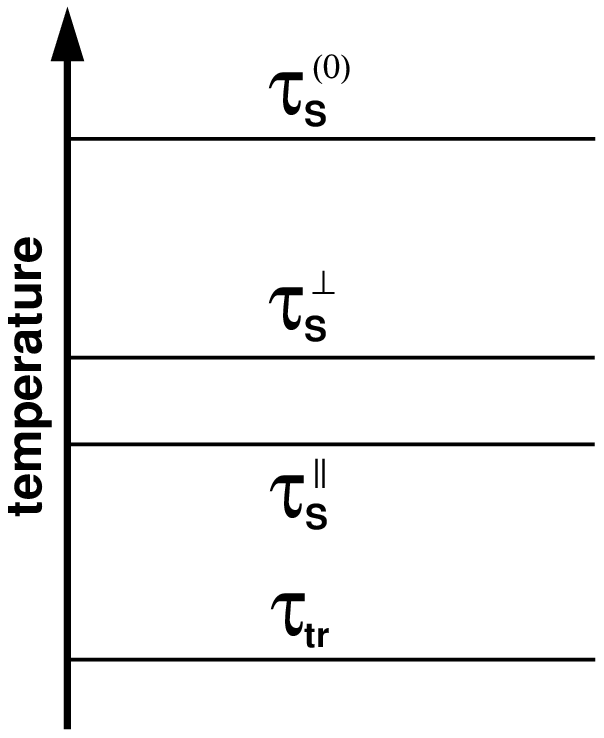,width=3cm}} 
\end{picture}
\caption{ Region of stability of the hexagonal phase in high shear
flow } 
\label{fig_diagram}
\end{figure}

Now we summarize our results.  In the high-shear limit the hexagonal
phase is found to be stable in the temperature range from $\tau_s$ to
$\tau_{tr}$ (see Fig.\ref{fig_diagram}).  Here we suppose that
hexagonal phase is the first phase appearing at cooling down from
melt.  If this is not a case, the temperature range
$(\tau_s,\tau_{tr})$ transforms to $(\tau_*,\tau_{tr})$, where
$\tau_*$ is a transition temperature from a hypothetic (bcc, gyroid,
...) to the hexagonal phase. The factor in front of $\cos{6\phi}$ in
the equation for $\tau_s^{(3)}$ is negative. This means that the
spinodal temperature is higher for the orientation with
$\phi=\pi/6$.  Moreover, the free-energy is minimal for this
orientation for all values of temperature from the range
$(\tau_s,\tau_{tr})$. Thus we predict the perpendicular orientation to
be the only stable orientation in the high-shear limit. 

We emphasize that the appearance of the angle-dependence in the free
energy only in the $O(\beta^{3}{\frak D}^3)$ is not a coincidence. It
reflects the internal symmetry of the system. Any figure on plane can
be oriented with respect to the particular direction only if its shape
deviates from circle.  In other words, if one expands the figure's
shape into the plane waves around circle:
$r(\phi)=R_0[1+\sum\alpha_{n}e^{in\phi}]$, the interaction with a
selected direction will appear in the first non-zero order.  For
hexagons it gives $n=3$. The n-th order in the expansion corresponds
to the interaction between $n$ different wavevectors.  Taking into
account that the parameter $\beta$ introduces the {\it pare-wise}
interaction between vectors (see (\ref{e2.8.1})), we conclude that the
angle-dependent terms in the free energy should be at least
$O(\beta^3)$. The role of the parameter $\frak D$ is different. Shear
breaks the rotational symmetry in the system, stipulating the
preferred (gradient) direction and allowing for a discrimination
between orientations.  Thus the angle-dependent terms in the free
energy should be proportional at least to the lowest possible power of
$\frak D$. This lowest possible power is determined by the concrete
form of the structure factor $S({\bv{k}})$. In (\ref{sigma_high})
$\beta$ and $\frak D$ enter as a one combination. So, one would expect
the free energy to be angle-dependent starting from the
$O(\beta^3{\frak D}^3)$ order. 

We illustrate the concept with the following example in which we
calculate the free energy of systems with other rotational
symmetry.  We consider for the moment an artificial phase which the
average density is given by: 
\begin{equation}
C({\bv{k}})=\sum_{i=1}^2{a_i(\delta_{{\bv{k}},q_0
{\bv{n}}^{(i)}}+\delta_{{\bv{k}},-q_0 {\bv{n}}^{(i)}})} 
\end{equation}
where ${\bv{n}}_i$ are two vectors lying in the plane perpendicular to
the velocity direction,
${\bv{n}}^{(1)}\cdot{\bv{n}}^{(2)}=\cos{\theta}$.  We consider two
cases: a) $\theta=\pi/2$, which describes square packed cylinders and
b) $\theta=\pi/12$, which describes highly non-symmetric cylindrical
pattern.  Calculating the structure constants (\ref{str_const}) and
integrating (\ref{e2.12}) with the help of (\ref{sigma_high}), we
obtain for the free energies: 
\begin{equation}
\label{f_0}
\Phi_{\theta=\pi/2}=-\frac{2\tau^2}{3\lambda}-
\frac{4I_1\tau\alpha^{2/3}}{3\lambda^{1/3}}{\frak
D}-\left[2I_1-\frac{3}{4}(I_3-I_2)^2\beta^2\cos{4\phi}\right]
\frac{\alpha^{4/3}\lambda^{1/3}}{3}{\frak
D}^2 
\end{equation}
\begin{eqnarray}
\label{f_15}
\Phi_{\theta=\pi/12}&=&-\frac{2\tau^2}{3\lambda}+
\frac{\tau\alpha^{2/3}}{3\lambda^{1/3}}{\frak
D}\left\{-4I_1+\frac{}{}\right. \nonumber \\
& &\left.\frac{\beta}{6}\left[-4(4+\sqrt{3})I_1+12(I_2+I_3)+
3(I_2-I_3)\left((2+\sqrt{3})\cos{2\phi}-\sin{2\phi}\right)\right]
\right\} 
\end{eqnarray}
In the case a) the structure deviates from a circle in the second
order ($\alpha_0=\alpha_1=0\,,\;\alpha_2\neq0$), while in the case b)
already in the first order. This gives $\beta^2$ and $\beta$ standing
in (\ref{f_0}) and (\ref{f_15}) in front of the angle-dependent
terms. Both cases have the same power of $\frak D$ as $\beta$ because
of (\ref{sigma_high}).  In the next section we shall see how a
different expression for the fluctuation integral will produce the
different lowest possible power of $D$. 

\section{Weak-shear behaviour}
\setcounter{equation}{0} In this section we consider the other limit
$D\rightarrow0$.  In this case the solution of
(\ref{e2.11})-(\ref{e2.12}) only slightly deviates from the
equilibrium one and we can construct a perturbation theory with a
small parameter $D$. Thus the structure factor is given by the
equation: 
\begin{equation}
\label{st_f}
S({\bv{p}})=\sum_{n=0}^\infty \left(\frac{D}{2\mu} \right)^n
S^{(n)}({\bv{p}})\, , \; S^{(n)}({\bv{p}})=\left[p_x
S_0({\bv{p}})\frac{\partial}{\partial p_y} \right]^n S_0({\bv{p}}) 
\end{equation}
where $S_0({\bv{p}})$ is the equilibrium structure factor. 

Following the conclusions of the previous section we evaluate
$S({\bv{p}})$ up to $O(\beta^3)$ and the lowest possible order in $D$,
which is $O(D)$. However, $S^{(1)}({\bv{p}})$ does not contribute to
the fluctuation integral.  Therefore, we keep $S^{(2)}({\bv{p}})$ in
the expression for the structure factor. Performing integration in
(\ref{e2.10}) with $S({\bv{p}})$ given by (\ref{st_f}) and using
(\ref{e2.10.2}) for $S_0({\bv{p}})$ we obtain the fluctuation
integral: 

\begin{equation}
\label{sigma_low}
\sigma(\hat{{\bv{p}}})=\sigma_{eq}(\hat{{\bv{p}}})-\frac{\pi
(\alpha\lambda)^3}{24\,r^{7/2}}\left(\frac{D}{D^*}\right)^2 \left\{1+
b_1 + b_2+\cdots\right\} \, , \; b_i\propto O(\beta^i) 
\end{equation}
where 
\begin{eqnarray}
b_1&=&\frac{1}{2r}(3e_{xx}+3e_{yy}+e_{zz})+\beta\frac{2\hat{p}_z^2-3}{7}
\nonumber \\ b_2&=&\frac{3}{8r^2}\left[\frac{}{}2
e_{ij}e_{ji}+e_{ii}(3e_{xx}+3e_{yy}-e_{zz})\right]-\nonumber \\
& &\frac{\beta}{3r}\left[\hat{p}_i
e_{ij}
\hat{p}_j+\frac{1}{2}e_{ii}+(1-\hat{p}_z^2)(e_{xx}+e_{yy})-
e_{zz}\hat{p}_z^2
\right] \nonumber 
\end{eqnarray}
$$
\sigma_{eq}(\hat{{\bv{p}}})=\frac{\alpha\lambda}{\sqrt{r}}
\left[1-\frac{1}{3}\beta+\frac{e_{ii}}{6r}
\right]+\frac{\alpha\lambda}{10r^{3/2}}\left[\frac{1}{4r}
(e_{ii}^2+2e_{ij}e_{ji})-\frac{\beta}{3}(e_{ii}+2\hat{p}_ie_{ij}
\hat{p}_j)
\right]+\cdots 
$$
and $\sigma_{eq}$ is an extension of (\ref{sigma_eq}) to the higher
orders in $\beta$. Here we assumed the summation over repeated
indices. 

In the equation (\ref{sigma_low}) and in what follows the terms of
$O(\beta^3)$ are cumbersome and we do not present them here. However,
we did use them in our calculations. 

The $\bv{p}$-independent terms in (4.2) contribute to the equation for
$r$: 
\begin{equation}
\label{r_low}
r=\tau+\lambda
A+\frac{\alpha\lambda}{\sqrt{r}}R_1-\left(\frac{D}{D^*}\right)^2
\frac{(\alpha\lambda)^3\pi}{24 r^{7/2}}R_2 
\end{equation}
where 
\begin{eqnarray}
&
&R_1=1+\frac{e_{ii}}{6r}+\frac{1}{40r^2}\left(e_{ii}^2+
2e_{ij}e_{ji}\right)+\cdots
\nonumber \\ R_2=1&+&\frac{3e_{xx}+3e_{yy}+e_{zz}}{2r} +
3\frac{e_{ii}(3e_{xx}+3e_{yy}-e_{zz})+2e_{ij}e_{ji}}{8r^2}+
\cdots\nonumber
\end{eqnarray}

For the anisotropy tensor (\ref{sigma_low}) together with
(\ref{e2.10.1}) gives a closed set of equations.  Iterating them up to
$O(D^2)$ and $O(\beta^3)$ we obtain: 
\begin{equation}
\label{e_low}
e_{ij}=E^{(1)}_{ij}-\left(\frac{D}{D^*}\right)^2
\frac{(\alpha\lambda)^3\pi}{24r^{7/2}}
E^{(2)}_{ij}+\cdots 
\end{equation}
where 
\begin{eqnarray}
E^{(1)}_{ij}&=&\beta\left\{\frac{\alpha\lambda}{3\sqrt{r}}\delta_{ij}+
\lambda\sum_{s=1}^3{a_s^2n^{(s)}_i
n_j^{(s)}}\right\}+\nonumber \\
& &\beta^2\left\{\frac{(\alpha\lambda)^2}{18r^2}
\delta_{ij}+\frac{\alpha\lambda^2}{30r^{3/2}}\sum_{s=1}^3
a_s^2\left[\delta_{ij}+2n^{(s)}_in^{(s)}_j(2-\delta_{ij})\right]
\right\}+\cdots
\nonumber \\
E^{(2)}_{ij}&=&\frac{\beta}{7}\left\{3\delta_{ix}\delta_{jx}+
3\delta_{iy}\delta_{jy}+\delta_{iz}\delta_{jz}
\right\}+\beta^2\left\{\frac{\alpha\lambda}{105r^{3/2}}
\left[59(\delta_{ix}\delta_{jx}+\delta_{iy}\delta_{jy})+22
\delta_{iz}\delta_{jz}\right]+\right.\nonumber
\\ &
&\left.\frac{\lambda}{6r}\sum_{s=1}^3a_s^2\left[\delta_{ij}+
2n^{(s)2}_y(\delta_{ix}\delta_{jx}+2\delta_{iy}\delta_{jy})+
4n^{(s)}_in^{(s)}_j
(1-\delta_{ij}) \right] \right\} + \cdots \nonumber 
\end{eqnarray}

Equations (\ref{r_low}) and (\ref{e_low}) are enough to construct the
free energy $\Phi$ as a solution of (\ref{pot}), where $h_i$, $r$ and
$e_{ij}$ are given by (\ref{e2.12}), (\ref{r_low}) and (\ref{e_low}),
respectively. Integration of (\ref{pot}) using (\ref{r_composite})
leads to the following equation for the free energy: 
\begin{equation}
\label{fe_low}
\Phi=F-\beta^3 \left(\frac{D}{D^*}\right)^2 {\frak a}^4 \frac{3\pi
\alpha^4 \lambda^5}{1280 r_0^6}\cos{6\phi} 
\end{equation}
where $\frak a$ is the equilibrium amplitude given by the equations
(\ref{a_eq}) and (\ref{r_equation}).  In (\ref{fe_low}) $F$ stands for
the angle-independent part of the free energy. 

The spinodal temperature $\tau_s$ is given by: 
\begin{equation}
\label{ts_low}
\tau_{s}=t+\beta^3 \left(\frac{D}{D^*}\right)^2 {\frak a}^4 \frac{\pi
\alpha^4 \lambda^6}{640 r_0^7}\cos{6\phi} 
\end{equation}
where $t$ includes the angle-independent terms. The spinodal
temperature is maximal for $\phi=0$. At the same time, the free energy
is minimal for the same orientation. It means that only the parallel
orientation is stable under low shear. 

To finish our symmetry analysis we note that in principle the
fluctuation integral (\ref{sigma_low}) contains all possible
combinations of powers of $\beta$ and $D^2$. According to the symmetry
arguments the angle-dependence appears in the lowest possible order in
$D$ (which is $D^2$ in this case) and the third order in $\beta$. This
is in agreement with (\ref{fe_low}) and (\ref{ts_low}). 

\section{Conclusions}
\setcounter{equation}{0} 

In the present paper we have shown how the orientational behaviour of
the hexagonal phase under simple shear flow can be described in the
framework of the dynamical model first developed by Fredrickson for
the lamellae \cite{main}. In this model the angle-dependence of the
4th order vertex function $\lambda$ (see \ref{e2.8.1}) plays a crucial
role, although it is of no importance in equilibrium (see Section II
for details). The parameter $\beta$ introduces the interaction between
structure and shear flow and allows for the rotational symmetry
breaking. The character of the interaction depends on the shear rate
and the resulting phase diagram has a complex structure. We predict
the parallel orientation to be stable at low and the perpendicular
orientation at high shear rates. Our results are in agreement with the
experimental dynamical phase diagram (Fig.\ref{fig2}) for the PE-PEP
system \cite{Tepe}. Our analysis shows that experimental study of the
hexagonal phase orientation is actually very difficult to
perform. Because of the symmetry arguments, the difference in the free
energies of the different orientations
$\Delta\Phi=\Phi_{\perp}-\Phi_{\parallel}$ is proportional to 
$$
\Delta\Phi=\left\{ 
\begin{array}{ll}
\beta^3 \frac{D^*}{D\;\,}&\!,\;D\rightarrow\infty \\ \\ \beta^3
\left(\frac{{D\;\,}}{{D^*}}\right)^2&\!,\;D\rightarrow0 \end{array}
\right. 
$$
This difference is extremely small (compare with
$\Delta\Phi\propto\beta$ for lamellae \cite{main}) and a particular
orientation is only slightly more stable than the other one. Some
crude experimental by-effects (like the ordering influence of the
sample walls) can easily suppress the pure orientational behaviour we
discussed. 

In comparison with the work \cite{main} we ignored the hydrodynamic
corrections to the effective shear rate $D_{\text{eff}}$ (see
Eq.(2.27) in \cite{main}).  These corrections are responsible for the
$\perp\rightarrow\parallel$ transition in the lamellar phase under
strong shear. Taken into account they would probably lead to the same
transition in the hexagonal phase. However, this is not an easy task
because one should keep terms up to $O(\beta^{3}{\frak D}^3)$ in order
to have an angle-dependent value. 

Another remaining problem concerns the stability of transversal
cylinders, i.e. cylinders with their axis oriented not in the
direction of flow \cite{report}. Our model cannot describe stability
of the transversal orientation \cite{C_M}.  Another approach should be
developed to complete our study. 

\section*{Acknowledgments}

We gratefully acknowledge Glenn~Fredrickson for explaining the details
of his work \cite{main}. The luminous discussions with A.~V.~Zatovsky
and V.~M.~Adamjan helped us to understand the role of hydrodynamics
and develop the theory group approach presented in Sections II and
III.  We also want to thank D.~Bedeaux and V.~L.~Koulinskii for the
general remarks on the subject, and Agur~Sevink for the friendly
atmosphere in our group. 

Our special thanks are to Nicolay Malomuzh for his enormous influence
and kind advice.

\begin{thebibliography}{10}
\bibitem{my_note} The only role of the dissipation in our model is to
renormalize the shear rate $D\,\rightarrow\,D_{\text{eff}}$. Since we
neglected any alteration of the velocity profile in the sample and
simply approximated $D_{\text{eff}}\approx D$, the existence of the
potential $\Phi$ is not surprising. Moreover, $D_{\text{eff}}$ is a
functional of the dissipative effects \cite{main}, so that only the
number $D_{\text{eff}}$ enters the dynamic equations. Thus, even in
this case the equations for the amplitudes $a_i$ have a potential
form. However, for the dissipative case it is much more difficult to
speculate about physical meaning of $\Phi$.
\end{thebibliography}
\end{document}